\documentclass[12pt,a4paper]{article}%
\usepackage{amssymb}
\usepackage{amsmath}
\usepackage{amsfonts}
\usepackage{graphicx}%
\providecommand{\U}[1]{\protect\rule{.1in}{.1in}}

\begin{document}

\title{Signed Sequential Rank CUSUMs }
\author{F. Lombard and C. van Zyl\\Centre for Business Mathematics and Informatics \\North-West University, Potchefstroom 2520, South Africa}
\date{}
\maketitle

\begin{abstract}
CUSUMs based on the signed sequential ranks\ of observations are developed for
detecting location and scale changes in symmetric distributions. The CUSUMs
are distribution free and fully self-starting: given a specified in-control
median and nominal in-control average run length, no parametric specification
of the underlying distribution is required in order to find the correct
control limits. If the underlying distribution is normal with unknown
variance, a CUSUM based on the Van der Waerden signed rank score produces
out-of-control average run lengths that are commensurate with those produced
by the standard CUSUM for a normal distribution with known variance. For
heavier tailed distributions, use of a CUSUM based on the Wilcoxon signed rank
score is indicated. The methodology is illustrated by application to real data
from an industrial environment.\bigskip\ 

\bigskip, symmetric distributions

\newpage

\end{abstract}

\section{Introduction}

\label{Introduction}

CUSUM procedures were developed to signal the onset of a persistent change
away from a specified product quality characteristic, such as a mean or
median. In the application treated in Section \ref{Application} of the paper,
the raw data consist of a series of matched pairs $(V_{1i},V_{2i}),\ i\geq1$,
the result from two treatments applied to the same sample of material. The
process is deemed to be in control as long as the treatment effect, defined as
the mean or median of $X_{i}=V_{1i}-V_{2i}$, is zero. Otherwise the process is
out of control. Page (1954) developed the first CUSUM for detecting a shift in
the mean of a normal distribution with known variance and in-control mean
zero. However, the normality assumption is often in doubt and it is known that
a standard normal CUSUM performs poorly when the true underlying distribution
has substantially heavier tails than a normal distribution (Hawkins and
Olwell, 1997, Section 3.7.1). It is therefore surprising that extension of the
methodology to heavier tailed, symmetric distributions has received almost no
attention in the literature. Furthermore, implementation of a normal
distribution CUSUM requires a known standard deviation, $\sigma$,$\ $in the
underlying distribution. Misestimation of $\sigma$ from Phase I data and
subsequent use of the estimate in the Phase II CUSUM can result in an
in-control average run length substantially different from the nominal value -
see, for instance, Hawkins and Olwell (1997, pages 159-161) and Keefe, et al.
(2015). This paper proposes CUSUM schemes that largely overcome these problems.

A natural approach towards extending a normal distribution CUSUM to other
symmetric distributions is to replace the observed data by rank-based
equivalents, which leads to distribution-free procedures. By "distribution
free" is meant that the in-control properties of the CUSUM do not depend on
the functional form of the underlying distribution. This paper develops
distribution-free CUSUMs for single observations, the SSR (signed sequential
rank) CUSUMs. The CUSUM is based on the series of signed sequential ranks
$s_{i}r_{i}^{+},\ i\geq1$, of the observations, where $s_{i}=sign(X_{i})$ and
$r_{i}^{+}$ is the rank of $|X_{i}|$ in the sequence $|X_{1}|,\ldots,|X_{i}|$,
that is, the number among $|X_{1}|,\ldots,|X_{i}|$ less than or equal to
$|X_{i}|$. When the process is in control, the $r_{i}^{+}$ form a series of
independent random variables with $r_{i}^{+}$ uniformly distributed on the
integers $1,\ldots,i$ - see Barndorff-Nielsen (1963, Theorem 1.1). Since the
$s_{i}$ are mutually independent and independent of the $|X_{i}|$ series,
$s_{i}r_{i}^{+},\ i\geq1$ is a sequence of independent random variables with
$s_{i}r_{i}^{+}$ uniformly distributed on $\pm1,\ldots,\pm i$, no matter what
the common distribution of the $X_{i}$ is. The independence, distribution
freeness and naturally sequential nature of the signed sequential ranks makes
them ideally suited to the construction of CUSUMs for time ordered data.
Control limits guaranteeing any specified in-control average run length (ARL)
can be determined once and for all for a range of reference constants and
nominal in-control average run lengths. The SSR CUSUM thus overcomes the
estimation problem that besets the standard normal CUSUM. Furthermore, its
validity is not dependent upon the existence of any moments in the underlying distribution.

Among other existing distribution-free CUSUMs for singly arriving observations
are the sequential rank CUSUM of McDonald (1990) and the changepoint CUSUM of
Hawkins and Deng (2010). These are based on unsigned (sequential or ordinary)
ranks of the $X$-data. Since unsigned ranks remain unchanged when the data are
transformed monotonically, these CUSUMs cannot incorporate specific
information about the in-control mean or symmetry of the in-control
distribution. As a result, they are not competitive with SSR CUSUMs when such
information is available. In fact, they are tailor made for situations in
which the in-control value of the characteristic in question, as well as the
underlying distribution, is unknown and where the current state of the
process, whatever it may be, is declared to be the in-control state. SSR
CUSUMs are not applicable in such instances.

The paper is structured as follows. The SSR CUSUMs are defined in Section
\ref{Signed SR cusum}. In Section \ref{OOC properties}, various out-of-control
properties of the CUSUMs are examined using theoretical results coupled with
Monte Carlo simulations. It is shown that the quantitative out-of-control
behaviour of an SSR CUSUM, which requires no knowledge of any parameters in
the underlying distribution, can be inferred from the behaviour of a standard
normal distribution CUSUM. In Section \ref{Selfstart} it is shown that a
distribution-free CUSUM based on the Van der Waerden rank score behaves like a
standard normal CUSUM, hence provides a simple solution to the unknown
variance problem that besets the latter. Section \ref{Omnibus} demonstrates
that a CUSUM based on Wilcoxon signed rank scores can serve usefully in an
omnibus role, especially when the underlying distribution has heavier tails
than the normal distribution. In Section \ref{Dispersion}, a sequential rank
CUSUM to detect a change in the dispersion of a symmetric distribution is
developed. Implementation of the proposed methodology is illustrated in
Section \ref{Application} by application to a set of data from a coal mining
operation. Section \ref{Summary} gives a summary of the main results and
conclusions. The supplementary material to the paper includes tables of
control limits guaranteeing a specified in-control average run length for
three specific distribution free CUSUMs.

\section{Signed sequential rank CUSUMs}

\label{Signed SR cusum}

The following generic version of CUSUM methodology will be used. Let $\xi_{i}$
be a function of $X_{1},\ldots X_{i}$ for which $E[\xi_{i}]=0$ when the
process is in control (the data come from a symmetric distribution with zero
median) and $E[\xi_{i}]>0$ when it is out-of-control. A one-sided (upper)
CUSUM then consists in computing recursively the sequence $D_{i}^{+}%
,\ i\geq1,$ where
\begin{equation}
D_{0}^{+}=0;\ D_{i}^{+}=max\left[  0,D_{i-1}^{+}+\xi_{i}-\zeta^{+}\right]
,\ i\geq1 \label{Generic cusum}%
\end{equation}
and where $\zeta^{+}>0$, the reference value, is a positive constant. The
CUSUM signals a change as soon as $D_{i}^{+}$ exceeds a control limit
$h^{+}>0$, the interpretation being that a change from an in-control to an
out-of-control situation has possibly occurred somewhere along the observed
sequence $X_{1},\ldots,X_{i}$. The run length, $N$, is the index at which a
signal first occurs. Because the barrier at $0$ forces the CUSUM\ to be
non-negative, the CUSUM will eventually signal regardless of whether a change
has taken place (a valid signal) or not (a false signal). To compensate for
this unit type I error, the control limit $h^{+}>0$ is chosen to ensure that
the average in-control ARL (IC ARL) equals a pre-specified value, denoted by
$ARL_{0}$.

To control for downward shifts, a second sequence%
\[
D_{0}^{-}=0;\ D_{i}^{-}=\min\left[  0,D_{i-1}^{-}+\xi_{i}+\zeta^{-}\right]
,\ i\geq1
\]
with control limit $h^{-}<0$ is computed. The CUSUM signals a shift as soon as
either $D_{i}^{+}>h^{+}$ or $D_{i}^{-}<h^{-}$. If the median of a symmetric
distribution is being monitored, one has $\zeta^{+}=\zeta^{-}=\zeta$ and
$h^{+}=h^{-}=h$. It is customary to exhibit the pairs $(i,D_{i}^{+})$ and
$(i,D_{i}^{-})$ in a single $(x,y)$ plot together with horizontal lines at the
control limits $y=h^{+}$ and $y=h^{-}$ - see Figure 1 in Section
\ref{Application}. A "normal CUSUM" is the special case in which $\xi
_{i}=X_{i}$ and the $X_{i}$ have a normal($0,1$) distribution. This CUSUM has
been studied extensively and its properties are well known - see Hawkins and
Olwell (1997, Chapter 3).

Let the score function\emph{ }$J(u),\ -1<u<1$ be odd and square-integrable on
the interval $(-1,1)$ with $\int_{0}^{1}J^{2}(u)du=1$ and set
\[
v_{i}^{2}=\frac{1}{i}\sum\nolimits_{j=1}^{i}J^{2}\left(  \frac{j}{i+1}\right)
.
\]
Then, under the in-control regime, the signed sequential rank statistics
\begin{equation}
\xi_{i}=J\left(  \frac{s_{i}r_{i}^{+}}{i+1}\right)  /\nu_{i}=s_{i}J\left(
\frac{r_{i}^{+}}{i+1}\right)  /\nu_{i}, \label{signed rank score}%
\end{equation}
$i\geq1$, are independently distributed with zero means and unit variances.
The proposed SSR (signed sequential rank) CUSUM consists in using $\xi_{i}$ in
the one-sided procedure (\ref{Generic cusum}) or in its two-sided version. If
the median shifts from zero to a non-zero value, or if the distribution
becomes asymmetric, the expected value of $\xi_{i}$ ceases to be zero.
Consequently, the CUSUM should be effective in detecting a shift away from
zero as well as detecting the onset of substantial asymmetry in the underlying distribution.

A Wilcoxon SSR CUSUM, abbreviated to "$W$-CUSUM", is based upon the $W$-score
\begin{equation}
J_{W}(u)=\sqrt{3}u. \label{Wilcoxon score}%
\end{equation}
Here, $v_{i}^{2}=(2i+1)/(2(i+1))$, whence
\[
\xi_{i}=\sqrt{\frac{6}{(2i+1)(i+1)}}s_{i}r_{i}^{+}%
\]
is used in (\ref{Generic cusum}). The Wilcoxon score is well suited to
practical implementation because of its simple form. Another popular score is
$\Phi^{-1}((1+u)/2)$, the Van der Waerden score. The corresponding CUSUM will
be referred to as the "$VdW$-CUSUM".

The "distribution-free when in control" character of SSR CUSUMs allows fairly
precise estimation by Monte Carlo simulation of the IC ARL for any given score
function $J$, control limit $h$ and reference constant $\zeta$. Tables S1 and
S2 in the supplementary material give control limits for a matrix of
$(\zeta,ARL_{0})$ pairs for use with the $W$- and $VdW$-CUSUMs.

Regardless of the reference value actually used, the existence of Phase I data
is not a prerequisite for initiating an SSR CUSUM. Given a reference constant
$\zeta$, any specified $ARL_{0}$ is guaranteed upon use of the appropriate
$h$. Thus, the SSR CUSUM\ is fully self-starting in the sense defined by
Hawkins and Olwell (1997) and the between-practitioner variation, as defined
in Saleh, et al. (2016), is zero. The effects of using a "wrong" $\zeta$ will
become evident only in the out-of-control ARL properties of the CUSUM. These
effects will be discussed in the sections that follow.

\section{Out-of-control properties}

\label{OOC properties}

Denote by $\tau$ the point in time (the changepoint) at which the underlying
process shifts from an in-control to an out-of-control state. The efficacy of
a CUSUM can be judged by the out-of-control average run length (OOC ARL)
\begin{equation}
E[N-\tau|N\geq\tau], \label{OOC ARL}%
\end{equation}
the expected time-to-signal after onset of an out-of-control state,
conditional upon no signal occurring prior to\ its onset. Some general
insights into the OOC\ ARL behaviour of SSR CUSUMs can be gained by
restricting attention to nominally "small" shifts and "large" changepoints
$\tau$. This criterion is in line with the primary objective of CUSUM
methodology, which is to detect quickly relatively small persistent shifts.
Furthermore, the shift $\delta$ in the median is expressed in units of an
(unknown) underlying scale parameter, $\sigma$, which is typically a measure
of dispersion in the underlying distribution. The fact that the ranks of any
set of data are scale invariant actually necessitates such a specification.
Define%
\begin{equation}
\theta_{0}=E[f_{0}(Y)J^{\prime}\left(  2F_{0}(Y)-1\right)  ], \label{theta_0}%
\end{equation}
where $f_{0}$ and $F_{0}$ denote the pdf and cdf of $Y=X/\sigma$, and notice
that $\theta_{0}$ is functionally independent of $\sigma$. Then, for $i\geq1$,%
\begin{equation}
E[\xi_{\tau+i}]\approx\theta_{0}\delta\neq0, \label{OOC mean}%
\end{equation}
implying that the CUSUM will show a sustained upward ($\delta>0$) or downward
($\delta<0$) drift after the changepoint, resulting in a finite OOC ARL.

It follows that the larger $\theta_{0}$ is, the better. Table 1 shows the
values of $\theta_{0}$ for the $W$- and $VdW$- CUSUMs in Student
$t$-distributions with $\nu$ degrees of freedom, standardized to unit standard
deviation for $\nu\geq3$ and to unit inter-quartile range for $\nu=2$ and
$\nu=1$. The $t$-distributions are chosen as benchmarks because they exhibit a
range of tail thicknesses that would mimic most cases occurring in practice.
Inspection of Table 1 reveals that the $W$-CUSUM should be the preferred one
among the two, \textit{except} when the distribution is normal ($\nu=\infty$).

\begin{center}%
\begin{tabular}
[c]{c}%
$\text{Table }$1\\
\multicolumn{1}{l}{$\text{Values of }\theta_{0}\text{ for the }W\text{- and
}VdW$-CUSUMs}\\
\multicolumn{1}{l}{$\text{in Student }t_{\nu}\text{-distributions.}$}%
\end{tabular}

$%
\begin{tabular}
[c]{c|c|c|c|c|c|}\cline{2-6}
& \multicolumn{5}{|c|}{$\nu$}\\\cline{2-6}
& $\infty$ & $4$ & $3$ & $2$ & $1$\\\hline
\multicolumn{1}{|l|}{$W$-CUSUM} & 0.98 & 1.18 & 1.37 & 1.18 & 1.10\\\hline
\multicolumn{1}{|l|}{$VdW$-CUSUM} & 1.00 & 1.12 & 1.29 & 1.06 & 0.93\\\hline
\end{tabular}
\ \ $
\end{center}

Because the distribution of partial sums of the $\xi_{i}$ tend to normality,
there is an expectation that the SSR CUSUMs will share some of the good
properties of a normal CUSUM. Indeed, Proposition 1 in the Appendix suggests
the following heuristic:%

\begin{equation}%
\begin{tabular}
[c]{|l|}\hline
\emph{If }$\delta$ \emph{is "small" and a shift of size }$\delta\sigma$
\emph{occurs at a "large" }$n=\tau$\emph{, }\\
\emph{then an SSR CUSUM with reference value }$\zeta$\emph{ and control limit
}$h$\\
\emph{behaves approximately like a normal CUSUM with the same }\\
$\zeta$ and $h\ $\emph{when a shift of size }$\delta\theta_{0}$\emph{ occurs
at }$n=\tau$\emph{.}\\\hline
\end{tabular}
\ \label{heuristic}%
\end{equation}
Some implications of this heuristic will now be explored.

\subsection{Specification of a reference value}

\label{Ref value}

The optimal choice of reference constant to detect a target shift of size
$\delta_{1}$ in a standard normal distribution CUSUM is $\delta_{1}/2$ - see,
for instance, Bagshaw and Johnson (1974, Section 2). Thus, the heuristic
(\ref{heuristic}) suggests $\zeta=\theta_{0}\delta_{1}/2$ as an appropriate
reference value in an SSR CUSUM. The variation of $\theta_{0}$ values seen in
Table 1 is not substantial so that default values $\theta_{0}=1$ or
$\theta_{0}=1.3$ seem appropriate, depending on the anticipated tail thickness
of the underlying distribution.

An estimate of $\theta_{0}$ is useful when designing a CUSUM - see Section
\ref{OOC ARL_1}. Such an estimate can be made if some Phase I data
$V_{1},\ldots,V_{m}$ are available. For the Wilcoxon score, for instance,
(\ref{theta_0}) reduces to $\theta_{0}=\sqrt{12}E[f(Y)]$, which can be
estimated by%
\begin{equation}
\hat{\theta}_{0}=\sqrt{12}\frac{\hat{f}_{0}(V_{1}/\hat{\sigma})+\cdots+\hat
{f}_{0}(V_{m}/\hat{\sigma})}{m} \label{theta_est}%
\end{equation}
where $\hat{\sigma}$ is a location invariant and scale equivariant estimator
of $\sigma$ (such as a sample standard deviation or inter-quartile range) and
$\hat{f}_{0}$ is an estimator of the density $f_{0}$ based upon the
observations $Y_{i}=V_{i}/\hat{\sigma},\ 1\leq i\leq m$. The suggested
reference value for use in Phase II is then $\hat{\zeta}=\hat{\theta}%
_{0}\delta_{1}/2$. Use of the appropriate control limit $\hat{h}$ (read from
Table S1 or Table S2, for instance) then guarantees a Phase II IC ARL equal to
the nominal value. There is again no practitioner-to-practitioner IC ARL variation.

\subsection{Out of control ARL}

\label{OOC ARL_1}

The heuristic (\ref{heuristic}) suggests that approximations to the OOC ARL of
the SSR CUSUM can be found by pretending that the underlying distribution is
normal. Such approximations are useful in CUSUM design. The following example
illustrates this numerically. Let $\ \zeta,\ h$ and $\tau$ be given. Denote by
$\mathcal{W}(\delta)$ and $\mathcal{N}(\delta)$ respectively the ARL of a one
sided $W$-CUSUM and a normal CUSUM at a persistent mean shift $\delta>0$ which
starts at $n=\tau$. The two CUSUMs use the same $\zeta$ and $h$. An
implication of the heuristic (\ref{heuristic}) is that
\begin{equation}
\mathcal{W}(\delta)\approx\mathcal{N}(\theta_{0}\delta) \label{ARL_approx}%
\end{equation}
when $\delta$ is "small". To gauge the extent to which this approximation is
useful, data were generated from two underlying distributions, a standard
normal distribution and a heavier tailed $t_{3}$ distribution, both
standardized to unit variance. Various mean shifts $\delta$ were induced at
$\tau=50$. For the normal distribution the $(\zeta,h)$ pairs $(0.1,12.01)$ and
$(0.25,7.25)$ were used and for the $t_{3}$ distribution the pairs
$(0.15,9.86)$ and $(0.35,5.66)$, based on larger $\zeta$ values to allow for
tail heaviness, were used. The $h$ values, taken from Table S2 in the
supplementary material, guarantee a $W$-CUSUM $ARL_{0}=500$ in all four cases.

$\mathcal{W}(\delta)$ was estimated from $10,000$ Monte Carlo trials in each
of the two distributions (normal and $t_{3}$), the estimates serving as
nominal "true" values of $\mathcal{W}(\delta)$. If an analytic formula or
software for determining the exact value $\mathcal{N}(\theta_{0}\delta)$ were
available, the quality of the approximation (\ref{ARL_approx}) could now be
assessed directly. However, except for $\tau=0$ and $\tau\rightarrow\infty$,
these are not available for arbitrarily specified $\tau$. In their absence a
"Monte Carlo formula" can be used. This entails estimating $\mathcal{N}%
(\theta_{0}\delta)$ for $\theta_{0}=0.98$ and $\theta_{0}=1.37$ at each of the
shifts $\delta$ by $10,000$ (or more) Monte Carlo trials \textit{using normal
random numbers only}.

The first column in Table 2 shows the first three in a series of shifts
$\delta=0.125:0.125:1.5$ that were induced at $\tau=50$. The columns headed
$d_{\zeta}$ show the differences $\mathcal{W}(\delta)-\mathcal{N}(\theta
_{0}\delta)$ between the true and predicted OOC ARLs rounded to the nearest
integer, the subscript on $d$ indicating the reference constant. The third
entry in each column shows the \textit{maximal} difference over all
$\delta>0.25$. (An unabridged version, Table S2.1, is in the supplementary
material.) Clearly, the normal approximation is excellent at all shift sizes
that would typically be considered to be of practical relevance and would
certainly be useful for the purpose of CUSUM design. In the design phase,
given an estimate $\hat{\theta}_{0}$ of $\theta_{0}$, the "Monte Carlo
formula" with various values of $\delta,\zeta,h$ and $\tau$ as inputs\ will
yield corresponding outputs $\mathcal{N}(\hat{\theta}_{0}\delta)$. These
outputs are estimates of the unknown $W(\delta)$ and can be used to gauge the
likely Phase II behaviour of the CUSUM under various specifications of the
input parameters.\newpage

\begin{center}%
\begin{tabular}
[c]{c}%
Table 2\\
\multicolumn{1}{l}{$W$-CUSUM ARL approximations in normal and $t_{3}$}\\
\multicolumn{1}{l}{distributions. $ARL_{0}=500$; changepoint $\tau=50$.}%
\end{tabular}

$%
\begin{tabular}
[c]{c|c|c|c|c|}\cline{2-5}
& \multicolumn{2}{|c}{{\normalsize normal}} &
\multicolumn{2}{|c|}{{\normalsize t}$_{3}$}\\\hline
\multicolumn{1}{|c|}{${\normalsize \delta}$} & ${\normalsize d}_{0.10}$ &
${\normalsize d}_{0.25}$ & ${\normalsize d}_{0.15}$ & ${\normalsize d}_{0.35}%
$\\\hline
\multicolumn{1}{|c|}{{\normalsize 0.125}} & {\normalsize -1} & {\normalsize 0}
& {\normalsize -1} & {\normalsize 5}\\\hline
\multicolumn{1}{|c|}{{\normalsize 0.25}} & {\normalsize 1} & {\normalsize 0} &
{\normalsize 1} & {\normalsize 3}\\\hline
\multicolumn{1}{|c|}{%
$>$%
0.25} & {\normalsize 1} & {\normalsize 1} & {\normalsize 2} & {\normalsize 2}%
\\\hline
\end{tabular}
\ \ \ \ \ \ \ \ \ \ $
\end{center}

The same simulations were also run at $\tau=0$, an instance in which the
condition in the heuristic that $\tau$ must be large is violated. As expected,
there were consistent differences between the (estimated) true values and the
values predicted by the heuristic. In particular, the heuristic
\textit{underestimated} substantially the true OOC ARL values. (Table S2.2 in
the supplementary material gives a full set of results.) This is not
surprising because accumulation of a non-negligible number of $\xi_{i}$ in
(\ref{Generic cusum}) is required to effect approximate normality. However,
the results in Table 2 suggest that $\tau=50$ observations is already
sufficient for this purpose even if the underlying distribution, such as a
$t_{3}$, has considerably heavier tails than a normal distribution.

\subsection{Behaviour under asymmetry}

\label{Asymmetry robustness}

Since an SSR CUSUM is constructed on an assumption of symmetry in the
in-control distribution, it should have an ability to detect asymmetry. This
aspect of SSR CUSUM behaviour was assessed in a small simulation study. The
in-control distribution was a standard normal distribution. From $\tau=51$
onwards, data were generated from skew-normal distributions (Azzalini, 2005)
with zero mean, unit variance and skewness parameters $\lambda=1$ (lightly
skewed), $\lambda=3$ (moderately skewed) and $\lambda=5$ (heavily skewed). A
two-sided $W$-CUSUM with $ARL_{0}=$ $500$ was run and the OOC ARL at each
value of $\lambda$ was estimated from $10,000$ simulated data sets. Table 3
shows the estimates.

\begin{center}%
\begin{tabular}
[c]{c}%
Table 3\\
\multicolumn{1}{l}{OOC ARL of $W$-CUSUM in}\\
\multicolumn{1}{l}{skew-normal distributions.}%
\end{tabular}

$%
\begin{tabular}
[c]{c|c|c|c|}\cline{2-4}
& \multicolumn{3}{|c|}{$\zeta$}\\\cline{2-4}
& $0.05$ & $0.15$ & $0.25$\\\hline
\multicolumn{1}{|c|}{$\lambda=1$} & $388$ & $421$ & $464$\\\hline
\multicolumn{1}{|c|}{$\lambda=3$} & $113$ & $119$ & $149$\\\hline
\multicolumn{1}{|c|}{$\lambda=5$} & $84$ & $82$ & $101$\\\hline
\end{tabular}
\ \ \ \ $
\end{center}

The very large out-of-control ARLs at $\lambda=1$ indicate that the CUSUM is
unable to detect efficiently such a small degree of asymmetry, thus implying
some robustness in that respect. On the other hand, the results at $\lambda=3$
and $\lambda=5$ indicate an ability to detect substantial degrees of
asymmetry. Consequently, a signal from the CUSUM is \textit{not necessarily}
an indication that the mean or median has changed. The subsequent data
analysis should include an assessment of the possibility that the signal
resulted from the onset of asymmetry.

\section{An efficient self-starting CUSUM for a normal distribution}

\label{Selfstart}

Suppose the data come from a normal distribution with unknown standard
deviation $\sigma$. A naive approach consists in estimating $\sigma$ from
Phase I data and pretending in Phase II that the estimate, $\hat{\sigma}$, is
error free. It is well known that such an approach is defective because the
Phase II IC ARL could differ vastly from the nominal value - see Keefe, et al.
(2015), where further references can also be found. Saleh, et al. (2016)
propose to ameliorate the effect by estimating appropriate control limits for
use in Phase II via bootstrapping from Phase I data. If such a method is used,
control limits must be generated afresh whenever the CUSUM is applied to a new
data series. A "once and for all" table, such as Table S1 or Table S2, is out
of the question.

A result of Chernoff and Savage (1958, Theorem 3) states that if $X$ has
finite variance and $J(u)$ is the $VdW$-score $\Phi^{-1}((1+u)/2)$, then
$\theta_{0}$ in (\ref{theta_0}) satisfies $\theta_{0}\geq1$, the minimum value
$\theta_{0}=1$ being attained only if $X$ has a normal distribution. This
fact, in conjunction with the heuristic (\ref{heuristic}), suggests that the
$VdW$-CUSUM offers a fully self-starting procedure that requires no
bootstrapping or parameter estimation of any kind. Since no Phase I data are
required and the CUSUM is guaranteed to achieve the specified $ARL_{0}$, there
is no between-practitioner variation. The $VdW$-CUSUM is asymptotically
efficient: asymptotically in the sense that $ARL_{0}$ should be "large" and
the OOC target "small"; and efficient in the sense that under these conditions
the OOC ARLs should be equal to those of a normal CUSUM with the same IC ARL
and the same OOC target. The only further restriction is that $\tau$ must be "large".

To form some idea of what "large" and "small" would mean in the present
context, ARLs of the normal- and $VdW$-CUSUMs were estimated by Monte Carlo
simulation at $ARL_{0}=500$ with typical target OOC shifts $\delta_{1}=0.5$
and $\delta_{1}=1.0$. Shifts ranging from $\delta=0.25$ to $\delta=1.50$ were
induced at $\tau=0,\ 50$ and at $100$. Denote by $\mathcal{V}(\delta)$ and
$N(\delta)$ the respective ARLs of the $VdW$- and normal CUSUMs. Table 4.1
shows the differences
\begin{equation}
d_{\delta}=V(\delta)-N(\delta)\text{,} \label{J_V perform}%
\end{equation}
rounded up to the nearest integer, at the various shifts $\delta$. The
boldface entries are those where the shift is greater than or equal to the target.

The only instance in which the relevant differences could be called
substantial is at $\tau=0$, a setting that violates the "large $\tau$"
requirement. Results at $ARL_{0}=1,000$ (Table S4.1 in the supplementary
material) follow the same pattern: a substantial difference at $\tau=0$ but a
difference of only $1$ at $\tau=50$ and $\tau=100$. Overall, the results
suggest that $\tau\geq50$ and $\delta_{1}\leq1$ meet the respective
descriptions "large" and "small" whenever $ARL_{0}\geq500$.

\begin{center}%
\begin{tabular}
[c]{c}%
$\text{Table\ 4.1}$\\
\multicolumn{1}{l}{$d_{\delta}$ from (\ref{J_V perform}) at $ARL_{0}=500.$}%
\end{tabular}

\begin{tabular}
[c]{c|c|c|c|c|c|c|}\cline{2-7}
& $\tau=0$ &  & $\tau=50$ &  & \multicolumn{2}{|c|}{$\tau=100$}\\\hline
\multicolumn{1}{|c|}{$\delta$} & $\delta_{1}=0.5$ & $\delta_{1}=1.0$ &
$\delta_{1}=0.5$ & $\delta_{1}=1.0$ & $\delta_{1}=0.5$ & $\delta_{1}%
=1.0$\\\hline
\multicolumn{1}{|c|}{0.25} & 8 & 24 & 2 & 10 & 2 & 9\\\hline
\multicolumn{1}{|c|}{0.4} & 7 & 20 & 1 & 7 & 0 & 4\\\hline
\multicolumn{1}{|c|}{0.5} & \textbf{7} & 18 & \textbf{1} & 4 & \textbf{1} &
2\\\hline
\multicolumn{1}{|c|}{0.75} & \textbf{7} & 13 & \textbf{1} & 2 & \textbf{1} &
1\\\hline
\multicolumn{1}{|c|}{1.0} & \textbf{7} & \textbf{11} & \textbf{1} & \textbf{1}
& \textbf{1} & \textbf{1}\\\hline
\multicolumn{1}{|c|}{1.25} & \textbf{7} & \textbf{11} & \textbf{1} &
\textbf{1} & \textbf{1} & \textbf{1}\\\hline
\multicolumn{1}{|c|}{1.5} & \textbf{8} & \textbf{11} & \textbf{1} & \textbf{1}
& \textbf{1} & \textbf{1}\\\hline
\end{tabular}

\end{center}

It is interesting to see what transpires when $\delta_{1}$ is apparently "not
small", say $\delta_{1}=2$. Then $\zeta=1$ is an optimal choice and the
control limit $h=2.2$ ensures $ARL_{0}=500$ in the $VdW$-CUSUM. The
appropriate control limit for the normal CUSUM\ is $h=2.323$. Table 4.2 shows
the results for $\tau=50$ and $\tau=100$.

\begin{center}%
\begin{tabular}
[c]{c}%
$\text{Table\ 4.}$2\\
\multicolumn{1}{l}{$d_{\delta}$ from (\ref{J_V perform}) at $ARL_{0}=500$ and
$\delta_{1}=2.0.$}%
\end{tabular}

\begin{tabular}
[c]{c|c|c|c|c|c|c|c|}\cline{2-8}
& \multicolumn{7}{|c|}{$\delta$}\\\cline{2-8}
& 0.25 & 0.50 & 1.00 & 1.50 & 2.00 & 2.50 & 3.00\\\hline
\multicolumn{1}{|c|}{$\tau=50$} & 27 & 29 & 6 & 2 & \textbf{2} & \textbf{1} &
\textbf{1}\\\hline
\multicolumn{1}{|c|}{$\tau=100$} & 22 & 15 & 2 & 1 & \textbf{1} & \textbf{1} &
\textbf{1}\\\hline
\end{tabular}

\end{center}

Again, the differences are practically negligible at the larger shifts
$\delta\geq2.0$. Overall, the Van der Waerden CUSUM, which does not require
any knowledge of the unknown $\sigma$, is not in any substantive manner
inferior to a normal CUSUM, which requires a known $\sigma$.

\section{An omnibus self-starting CUSUM}

\label{Omnibus}

From Table 1 it is clear that $\theta_{0}$ is larger for the Wilcoxon score
than for the Van der Waerden score, except when the underlying distribution is
normal, the difference there being almost negligible. Thus, the $W$-CUSUM
would be preferred in distributions with heavier than normal tails and would
be almost as good as the $VdW$-CUSUM in a normal distribution. This accords
with the conclusions in Hodges and Lehmann (1960, Section 5) regarding the
relative performances of the two scores in a hypothesis testing context. The
$W$-CUSUM can therefore be recommended as an omnibus self-starting CUSUM that
will be effective in many situations. The following are some possible
limitations of the $W$-CUSUM.

First, the CUSUM has $|\xi_{i}|\leq\sqrt{3}$. Thus, if the target $\delta_{1}$
exceeds $2\times\sqrt{3}=6.92$ and the default reference constant
$\zeta=\delta_{1}/2$ is used, the ARL at all $\delta>0$ will be infinite
because then $\xi_{i}-\zeta$ is always negative, whence $D_{n}^{+}=0$ for all
$n$. However, this is not a substantive practical limitation because the
typical range of out-of-control target shifts $\delta_{1}$ in applications of
the CUSUM are considerably less than $6.92$.

Second, given $\zeta<2\sqrt{3}$ and $h$, a $W$-CUSUM requires at least
$[h/(\sqrt{3}-\zeta)]+1$ observations to reach the control limit. The maximum
value of this quantity over all $(\zeta,h)$ pairs in Table S1 is $5$
observations. A maximum of five possible additional observations seems a small
price to pay for the simplicity involved in applying the $W$-CUSUM and reaping
the benefits of (i) its high efficiency in non-normal distributions and of
(ii) its bounded score function, which inhibits transient special causes from
producing signals - see Section \ref{Application} for an example.

When observations occur naturally in groups of two or more without a time
ordering, no SSR CUSUM is applicable because sequential ranks are then not
uniquely defined. In such a case the grouped signed rank CUSUM of Bakir and
Reynolds (1979), which is also distribution-free, can be used.

\section{A sequential rank CUSUM for dispersion}

\label{Dispersion}

While the in-control properties of the SSR CUSUM do not depend upon the
variability of the underlying data, its proper application does require\ the
variability to remain unchanged. Suppose that after $\tau>0$ observations
there is a change to a distribution with density $g(x)=f(x/\sigma)/\sigma$,
$\sigma\neq1$. Then $\sigma$ is the fraction by which the current, unknown,
dispersion changes. To detect an increase in dispersion, one can use a CUSUM
based on the scores $J^{2}(u)$, thus eliminating the effect of the sign of
$X$. With the Wilcoxon score, the corresponding sequential rank statistic to
be used in (\ref{Generic cusum}) is then
\begin{equation}
\xi_{i}=\frac{6(r_{i}^{+})^{2}}{(2i+1)(i+1)}-1, \label{disp summand}%
\end{equation}
which has zero in-control expected value. The corresponding CUSUM will be
referred to as the "$W^{2}$-CUSUM". Since the sequential ranks $r_{i}^{+}$ are
invariant under scale changes, it is clear that a CUSUM based on them cannot
detect changes from a specified value of $\sigma$. Only changes away from the
current value of $\sigma$, whatever it may be, will be detectable, and this
only if the change occurs after a sufficiently long time lapse $\tau>2$.
Furthermore, the effect of the pre-change value of the scale parameter becomes
negligible as observations continue to accrue after a change to different
value. Thus, the CUSUM will eventually return to a nominally in-control state
after a change has occurred. This behaviour is similar to that of
self-starting CUSUMs, and is a warning to users of the need for corrective
action as soon as a change is diagnosed - see Hawkins and Olwell (1997,
Section 7.1).

If a change from an unspecified $\sigma$ to $\sigma\Delta,\ \Delta>0$, occurs,
the analogue of (\ref{OOC mean}) is,
\[
E[\xi_{\tau+i}]\approx\theta_{1}log\ \Delta
\]
where%
\[
\theta_{1}=12E\left[  \left(  2F_{0}(Y)-1\right)  Yf_{0}(Y)\right]  .
\]
Thus, appropriate reference constants for a $100\alpha\%$ change up or down
from the current dispersion level would be $\zeta^{+}=\theta_{1}%
log\ (1+\alpha)\ /2$ in an upper CUSUM and $\zeta^{-}=-\theta_{1}%
log\ \alpha\ /2$ in a lower CUSUM. Table 6 shows values of $\theta_{1}$ in
some $t_{\nu}$ distributions. It seems that $\theta_{1}=1$ could serve
usefully as a default value. If an estimator is required,
\[
\hat{\theta}_{1}=12\sum\nolimits_{i=1}^{m}\left(  2\frac{i}{m+1}-1\right)
\frac{V_{(i)}}{\hat{\sigma}}\hat{f}\left(  \frac{V_{(i)}}{\hat{\sigma}%
}\right)
\]
will do, where $V_{(1)}<\cdots<V_{(m)}$ denote the order statistics of the
Phase I data and where $\hat{f}$ and $\hat{\sigma}$ are as in (\ref{theta_est}%
). Table S6 in the supplementary material has control limits that cover all
$\theta_{1}$ values in Table 5 and all $0.5<\alpha<1$.

\begin{center}%
\begin{tabular}
[c]{c}%
$\text{Table }$5\\
\multicolumn{1}{l}{$\text{Values of }\theta_{1}\text{ for the }W^{2}%
\text{-CUSUM}$}\\
\multicolumn{1}{l}{$\text{in }t_{\nu}\text{-distributions.}$}%
\end{tabular}

$%
\begin{tabular}
[c]{c|c|c|c|c|c|}\cline{2-6}
& \multicolumn{5}{|c|}{$\nu$}\\\cline{2-6}
& $\infty$ & $4$ & $3$ & $2$ & $1$\\\hline
\multicolumn{1}{|c|}{$\theta_{1}$} & 1.10 & 0.94 & 0.89 & 0.8 & 1.10\\\hline
\end{tabular}
\ \ $
\end{center}

\section{Application}

\label{Application}

The data consist of successive pairs of determinations $(V_{1i},V_{2i})$,
$1\leq i\leq240$ of the ash content of coal, reported as a percentage per unit
mass, from two nominally identical laboratories. The measurements $V_{1i}$ and
$V_{2i}$ were made on two identical coal samples extracted from a single batch
of coal. If the true value of the ash content is $T_{i}$, then the
determinations by the two laboratories may be represented as
\[
V_{1i}=T_{i}+\varepsilon_{1i},\ V_{2i}=T_{i}+\varepsilon_{2i}%
\]
where $\varepsilon_{1i}$ and $\varepsilon_{2i}$ represent the respective
laboratory measurement errors. These errors may be taken to be statistically
independent since the laboratories operate independently of one another. Given
that the laboratories are operating to ISO or ASTM specifications, the errors
should also be identically distributed with zero means and common, albeit not
precisely known, standard deviation $\sigma$. The $T_{i}$ reflect the
characteristics of various seams from which the coal is extracted and are
typically neither independently nor identically distributed. Nevertheless, the
differences
\[
X_{i}=V_{1i}-V_{2i}=\varepsilon_{1i}-\varepsilon_{2i},
\]
which are the focus of interest here, do not depend upon the $T_{i}$ and are
independently and symmetrically distributed around zero\textbf{.
}{\normalsize A non-zero mean or asymmetry in the distribution, or a change in
the variance of }$X$, {\normalsize indicates a deviation from specifications
in one or both of the laboratories. This would typically lead to an audit of
the analysis procedures used in the laboratories to isolate the cause of the
deviation. }

{\normalsize To monitor the mean and standard deviation of }$X$, four CUSUMs
with $ARL_{0}=2,000$ were run concurrently: a two-sided $W$-CUSUM - see
(\ref{Wilcoxon score}) - for the mean and a two-sided $W^{2}$-CUSUM - see
(\ref{disp summand}) - for the standard deviation. The overall IC ARL would
then be approximately $500$. {\normalsize In the present instance, no formal
Phase I data were available. However, based on the operating specifications
for ash analysis, the measurement error }$\varepsilon$ in a laboratory should
have a standard deviation of about $0.45$ ($\%$ ash per unit mass of coal),
implying that {\normalsize the standard deviation, }$\sigma$, {\normalsize of
}$X$ should be between $0.6$ and $0.7$. {\normalsize The target mean change
size was specified as $\delta_{1}=0.25$}. To accommodate heavier than normal
tails, $\theta_{0}=1.18$ from Table 1 was used to arrive at a reference value
$\zeta=1.18\times0.25/2\approx0.15$ for the CUSUM. The corresponding control
limit from Table S1 {\normalsize is }$h=14.063$. {\normalsize To detect a
}$50\%$ increase or decrease {\normalsize in dispersion with the dispersion
CUSUM, reference values of }$\zeta^{+}={\normalsize 0.20}$ $(\approx
log\ 1.5\ /2)$ and $\zeta^{-}=-0.35\ (\approx log\ 0.5\ /2)$ are used.
{\normalsize The control limits }from Table S6 are $h^{+}=10.29$ and
$h^{-}=-6.29$.

The CUSUMs are shown in Figure 1. The $W$-CUSUM {\normalsize (left-hand panel)
signals an increase in the mean at observation $235$. The usual CUSUM-based
estimator of the changepoint after occurrence of a signal is the last index at
which the CUSUM (upper or lower) was at zero, which in this case is }%
$\hat{\tau}=214$. The locations of both on the time axis is indicated by
vertical dotted lines. The estimate of the new mean from observations $215$
through $235$ is $0.595$ while the mean of the first $214$ observations is
$-0.027$. The change in the mean from $-0.027$ to $0.595$ is highly
statistically significant ($p$-value $=0.0001$ from a bootstrap two sample
$t$-test on $10,000$ bootstrap samples). {\normalsize The time series plot in
Figure 2} shows an apparent outlier at $X_{103}=3.81$ which, after
investigation turned out to be due to a transcription error in a spreadsheet.
That this value was not detected by either of the CUSUMs points to their
robustness against transient special cause effects. Figure 3 shows a Q-Q plot
(left panel) and kernel density estimate (right panel) made from the data
$\{X_{1},\ldots,X_{214}\}$ after removal of the outlier. Both plots suggest a
degree of non-normality and slight asymmetry in the underlying distribution.
In view of the highly significant difference between the means, it is unlikely
that asymmetry was the cause of the CUSUM signal.

\begin{center}%
\[%
\begin{array}
[c]{cc}%
{\includegraphics[
height=2.4829in,
width=2.5183in
]%
{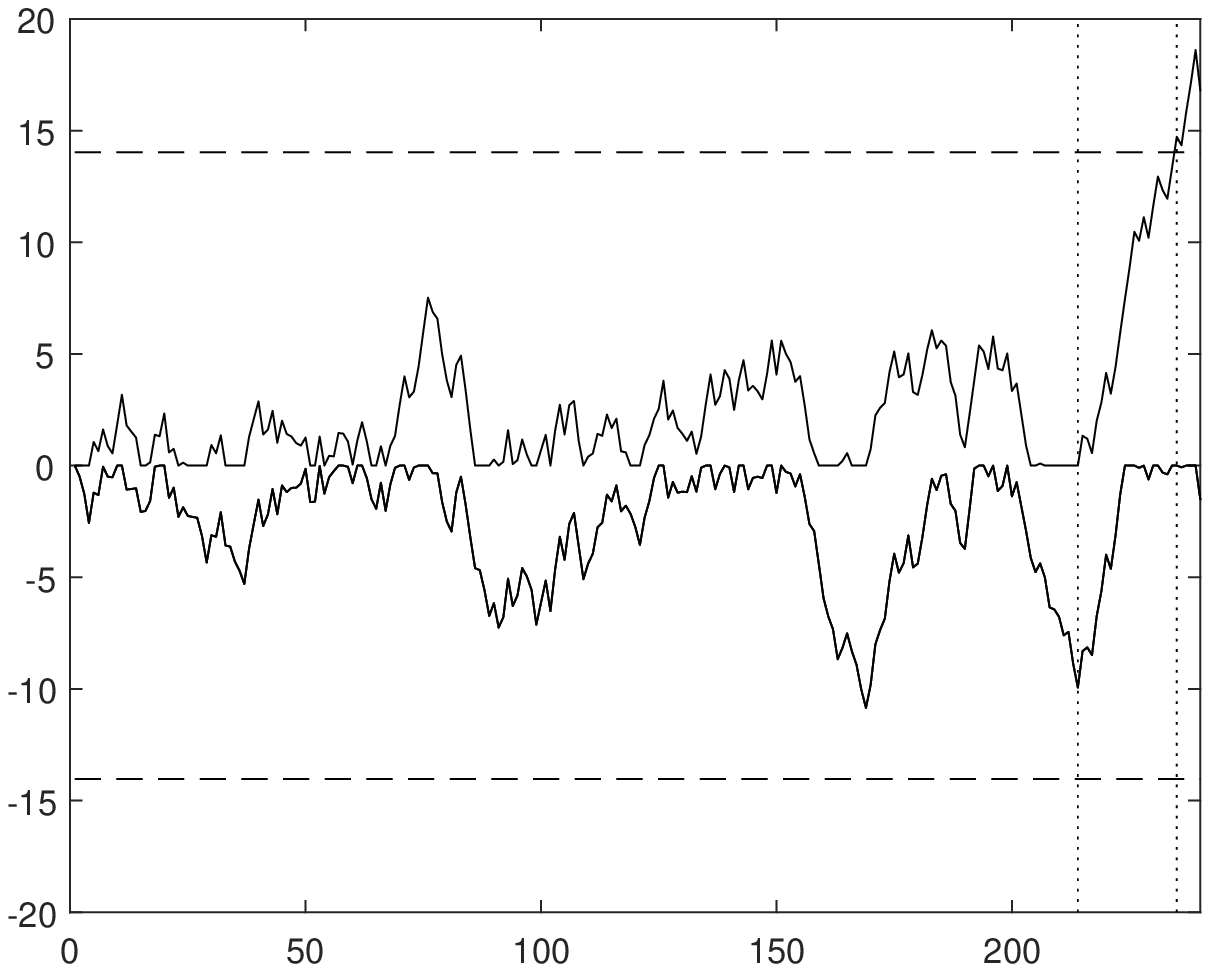}%
}
&
{\includegraphics[
height=2.4561in,
width=2.6852in
]%
{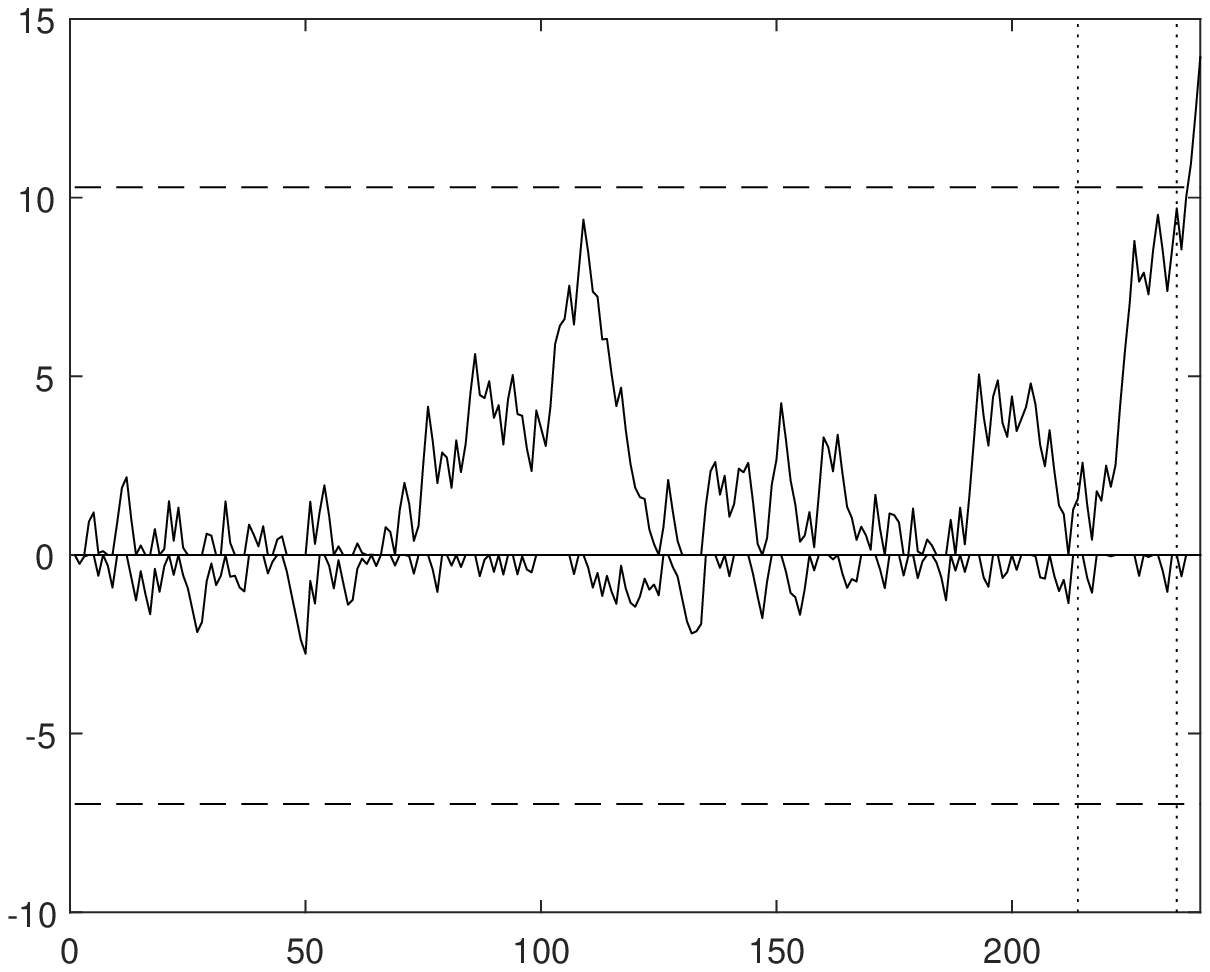}%
}
\end{array}
\]

$%
\begin{tabular}
[c]{l}%
Fig. 1. $W$-CUSUM (left panel) and $W^{2}$-CUSUM (right panel) CUSUMs of the\\
ash data. Observation index $n$ on the horizontal axis. Horizontal dashed\\
lines indicate the control limits.
\end{tabular}
\ \ $\newpage

$%
{\includegraphics[
height=2.4517in,
width=3.7896in
]%
{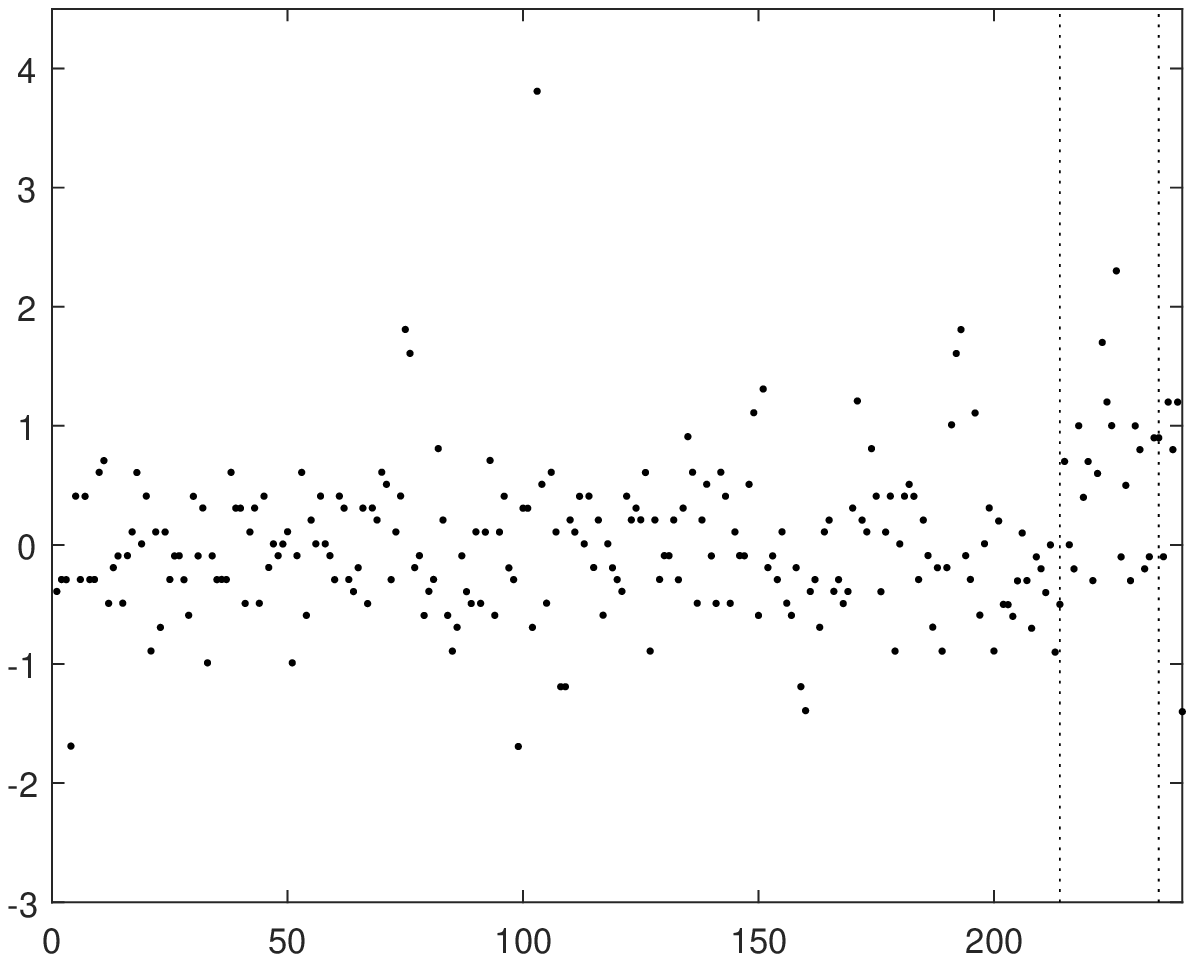}%
}
\ \ \ \ $%

\begin{tabular}
[c]{c}%
Fig. 2. Time series plot of dataset $X_{1},\ldots,X_{240}$.
\end{tabular}

\[%
\begin{array}
[c]{cc}%
\raisebox{-0.0277in}{\includegraphics[
height=2.2459in,
width=2.4777in
]%
{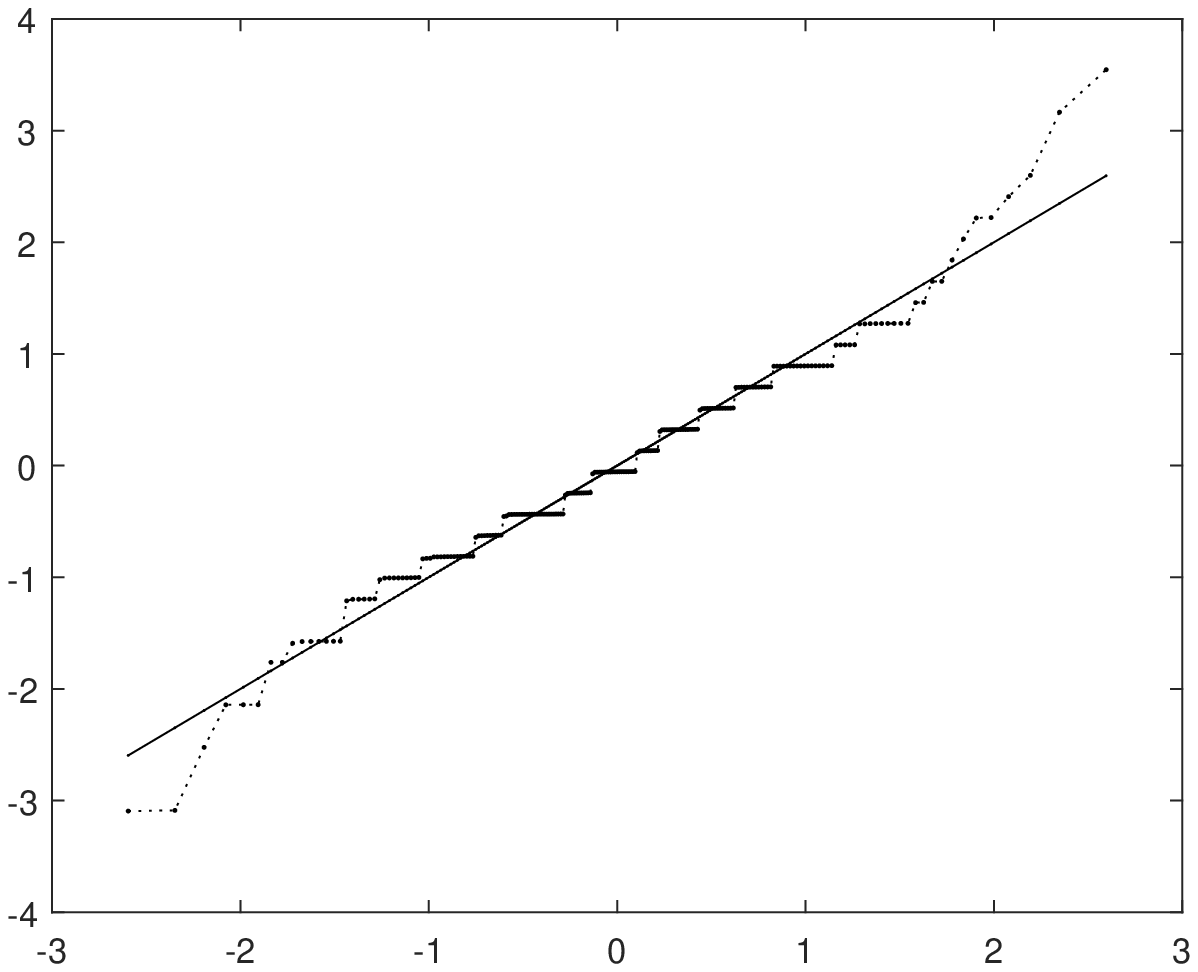}%
}
&
{\includegraphics[
height=2.2528in,
width=2.4984in
]%
{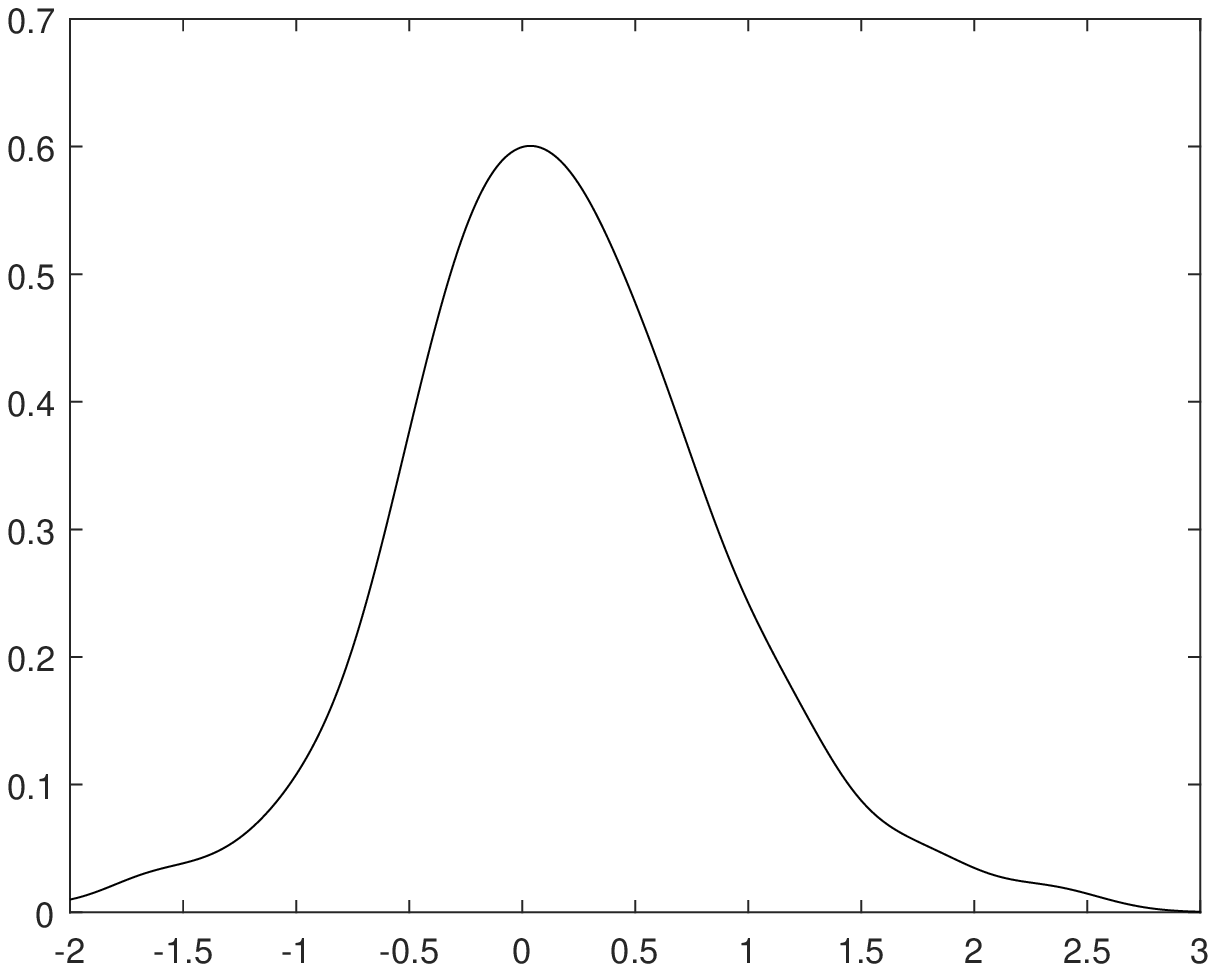}%
}
\end{array}
\]

$%
\begin{tabular}
[c]{l}%
Fig. 3. Q-Q plot (left panel) and density estimate (right panel)\\
from data $X_{1},\ldots,X_{214}$ after removal of one outlier.
\end{tabular}
\ $
\end{center}

The $J_{W}^{2}$ CUSUM (right-hand panel in Figure 1) signals at $n=238$,
shortly after the $W$-CUSUM. The standard deviation estimates from the
segments $\{X_{1},\ldots,X_{214}\}$ and $\{X_{215},\ldots,X_{235}\}$ are very
similar, namely $\hat{\sigma}_{1}=0.61$ and $\hat{\sigma}_{2}=0.68$. A
bootstrap $F$-test for equality of variances in these two segments yields a
$p$-value of $0.66$. Thus, on the available evidence, the signal from the
variance chart is most likely a result of the substantial mean change. Further
substantiation of this conclusion comes from a Monte Carlo simulation in which
data were generated from the density estimate in Figure 3, shifted to the
right by an amount $0.027$ to make the resulting density have mean zero.
Repeated sampling from this distribution ensures a constant standard
deviation. An increase of $0.622\ (=0.595+0.027)$ was induced in the median
after $\tau=214$ observations. The estimate of the ARL $E[N-\tau|N>\tau]$
resulting from $10,000$ such trials was $22$. This is of the same order of
magnitude as the excess $N-\hat{\tau}=235-214=21$ in the observed data and
confirms the likely reaction by the dispersion CUSUM to the mean shift.

The impact of the choices $\zeta=0.15$ and $\zeta=0.20$ on the CUSUMs can be
assessed if the first $50$ observations, say, are treated as in-control Phase
I data. These have a standard deviation of $\hat{\sigma}=0.45$, which is
somewhat less than the original estimate of $0.6$. The default bandwidth for a
Gaussian kernel density estimate made on these Phase I data is $b=0.22$. Then,
using (\ref{theta_est}), a computation gives $\hat{\theta}_{0}=1.03$ so that
the suggested reference constant for a target shift of $\delta_{1}=0.5\ $would
be
\[
\zeta=1.03\times0.5/2\approx0.25\text{.}%
\]
For the $W^{2}$-CUSUM, analogous computations give
\[
\hat{\theta}_{1}=1.12,
\]
which suggests $\zeta=0.23$ as reference constant, very close to the value
that was actually used. Running the CUSUMs on the Phase II observations
$X_{51},X_{52},\ldots$ with this new reference constant has no material effect
on the results: The $W$-CUSUM then signals at $n=50+180=230$ and the
changepoint is again estimated to be at $n=50+164=214$.

For $b=0.44$ and $b=0.11$, respectively double and one half the default
bandwidth, the corresponding estimated reference values for the $W$-CUSUM are
$\zeta=0.21$ and $\zeta=0.28$. Again, when these are used, the CUSUM results
are almost identical to those found at $\zeta=0.15$. This points to the fact
that the performance of the CUSUM is not overly sensitive to misestimation of
$\theta_{0}$ and consequent misestimation of the "optimal" reference value
$\zeta=\theta_{0}\delta_{1}$.

\section{Summary}

\label{Summary}

This paper develops CUSUMs based on signed sequential ranks to detect changes
away from a specified in-control median of an unknown symmetric distribution.
The in-control behaviour of the CUSUMs is distribution-free while their
out-of-control properties are shown to be well approximated by those of a
normal distribution CUSUM. In particular, no estimates of distribution
parameters are required to initiate the CUSUMs and they exhibit no
between-practitioner variation. When the underlying distribution is normal
with unknown variance, a CUSUM based on the Van der Waerden rank score is
efficient compared to a normal distribution CUSUM. A Wilcoxon-type CUSUM is
fully self-starting and near-efficient for heavy tailed distributions. A CUSUM
to detect changes in dispersion is also developed. The methodology is
illustrated\ in an application to a set of data from an industrial
environment.\bigskip

{\large Acknowledgement.} This work was supported by the National Research
Foundation of South Africa under grant number 96140

\section{Appendix}

\subsection{A justification for the heuristic (\ref{heuristic})}

The following result, which is a special case of Theorem 1 in Lombard (1981),
forms the basis of the heuristic and follows upon making identifications
between the notation used in this paper and that used in Lombard (1981).
Alternatively, the result can be obtained from Theorem 1 of Lombard and Mason
(1985) upon making appropriate Taylor expansions.\newpage

{\large Proposition 1}\bigskip

Let $h$ and $\tau^{\ast}$ be positive numbers. For every $t>0$, denote the
integer part of $h^{2}t$ by $\lfloor h^{2}t\rfloor$ and set $\tau=\lfloor
h^{2}\tau^{\ast}\rfloor$. Suppose the independent observations $X_{1}%
,\ldots,X_{\tau}$ have common cdf $F(x)$ while $X_{\tau+1},X_{\tau+2},\ldots$
have cdf $F(x-\delta\sigma/h)$. With $\xi_{i}$ from (\ref{signed rank score}),
set\textbf{ }%
\begin{equation}
S_{\lfloor h^{2}t\rfloor}=%
{\textstyle\sum\nolimits_{i=1}^{\lfloor h^{2}t\rfloor}}
(\xi_{i}-\zeta),\ t\geq0 \label{Cont time approx}%
\end{equation}
where $\zeta=\gamma/h$ and $S_{0}=0$. Then the continuous time process
$S_{\lfloor h^{2}t\rfloor}/h,\ t\geq0$ converges in distribution as
$h\rightarrow\infty$ to the continuous time process%
\begin{equation}
\Theta(t)=W(t)-\gamma t+\theta\delta\sigma max\{0,t-\tau^{\ast}\}
\label{cont time limit}%
\end{equation}
where $W$ denotes a standard Brownian motion and where
\[
\theta=-\int_{-1}^{1}J(u)\frac{f^{\prime}(F^{-1}(\frac{1+u}{2}))}%
{f(F^{-1}(\frac{1+u}{2}))}du\text{.}%
\]
Here, convergence in distribution is meant in the sense of weak convergence of
probability measures on the space $D[0,\infty)$ - see Billingsley (1999,
Section 16).$\bigskip$

Straightforward calculation involving an integration by parts gives
$\theta=\theta_{0}/\sigma$ for $\theta_{0}$ defined in (\ref{theta_0}). Then,
upon evaluating (\ref{Cont time approx}) and (\ref{cont time limit}) at
$t=n/h^{2},\ 1\leq n\leq\tau$ and $t=(\tau+k)/h^{2},\ k\geq1$, and using the
fact that $W(n/h^{2})$ and $W(n)/h,\ n\geq1,$ have the same joint
distributions, Proposition 1 \textit{suggests} that the joint distributions of
the partial sums $S_{n},\ n\geq m$ that figure in the SSR CUSUM can be
approximated by those of the sequence%
\[
\Theta(n)=W(n)-\zeta n+\theta_{0}\delta max\{0,n-\tau\}
\]
where $\tau>m$ and $m$ is a "large" positive integer. Let $\xi_{1}^{\ast
},\ldots\xi_{\tau}^{\ast}$ be i.i.d. normal$(0,1)$, let $\xi_{\tau+k}^{\ast
},\ k\geq1$ be i.i.d. normal($\theta_{0}\delta,1)$ and set $S_{n}^{\ast}%
=\xi_{1}^{\ast}+\cdots+\xi_{n}^{\ast}-n\zeta$. Then the sequences\emph{
}$S_{n}^{\ast},$\emph{ }$n\geq1$ and $\Theta(n),\ n\geq1$\emph{ }are
identically distributed. Thus, the distribution of the normal CUSUM based on
$\xi_{n}^{\ast},\ n\geq1$ provide an approximation to the SSR CUSUM based on
$\xi_{n}^{\ast},\ n\geq1$. This is the content of the heuristic.\newpage

\section{Supplementary Material}

\subsection{Control limits for the W- and VdW-CUSUMs (Section 2)}

The computations used to obtain the control limits in Tables S1 and S2 below
were as follows. Since the partial sums of the $\xi_{i}$ are approximately
normally distributed, it is not difficult to imagine that the control limits
$h$ of the CUSUM will correspond closely to those of a standard normal cusum.
Given a set of reference values and nominal IC ARL values $ARL_{0}$, denote by
$h_{1}$ the corresponding control limits from a standard normal CUSUM. The
first step of an iterative process was to estimate the IC ARL of the SSR CUSUM
on a $(\zeta,h_{1})$ grid using, for instance, $10,000$ independent Monte
Carlo generated realizations with a uniform distribution on $[-1,1]$ serving
as in-control distribution. Denote the estimate by $\hat{A}(\zeta,h_{1})$.
Cubic spline interpolation from $(\zeta,\hat{A}(\zeta,h_{1}))$ to $(\zeta,h)$
then yielded new estimates, $h_{2}$, of the correct control limits. A further
$10,000$ independent Monte Carlo generated realizations using $h_{2}$ produced
a new estimated IC ARL $\hat{A}(\zeta,h_{2})$. This process was repeated until
all the differences $|\hat{A}(\zeta,h)-ARL_{0}|$ were less than $3$. For
$\zeta\leq0.25$, no more that three iterations were required, while for
$\zeta\geq0.25$, six iterations sufficed. Finally, the control limits were all
checked independently in $100,000$ Monte Carlo runs. The largest difference
between nominal and simulation estimated IC ARLs was $3$.\bigskip

\begin{center}
$%
\begin{array}
[c]{c}%
\text{Table S1}\\
\multicolumn{1}{l}{\text{Control limits for the }W\text{ - CUSUM.}}%
\end{array}
$

$%
\begin{tabular}
[c]{c|c|c|c|c|c|}\cline{2-6}
& \multicolumn{5}{|c|}{$ARL_{0}$}\\\hline
\multicolumn{1}{|c|}{$\zeta$} & $100$ & $250$ & $500$ & $1,000$ &
$2,000$\\\hline
\multicolumn{1}{|c|}{$0.1$} & 6.45 & 9.44 & 12.01 & 14.79 & 17.93\\\hline
\multicolumn{1}{|c|}{$0.15$} & 5.65 & 7.91 & 9.86 & 11.88 & 14.06\\\hline
\multicolumn{1}{|c|}{$0.2$} & 5.00 & 6.89 & 8.37 & 9.96 & 11.57\\\hline
\multicolumn{1}{|c|}{$0.25$} & 4.46 & 6.02 & 7.25 & 8.52 & 9.84\\\hline
\multicolumn{1}{|c|}{$0.3$} & 4.01 & 5.33 & 6.37 & 7.45 & 8.53\\\hline
\multicolumn{1}{|c|}{$0.35$} & 3.62 & 4.75 & 5.66 & 6.58 & 7.51\\\hline
\multicolumn{1}{|c|}{$0.4$} & 3.29 & 4.29 & 5.06 & 5.87 & 6.66\\\hline
\multicolumn{1}{|c|}{$0.45$} & 2.99 & 3.89 & 4.56 & 5.24 & 5.96\\\hline
\multicolumn{1}{|c|}{$0.5$} & 2.73 & 3.52 & 4.13 & 4.74 & 5.34\\\hline
\end{tabular}
\ \ \ $\newpage

$%
\begin{array}
[c]{c}%
\text{Table S2}\\
\multicolumn{1}{l}{\text{Control limits for the }VdW\text{-CUSUM. For }%
ARL_{0}>1000}\\
\multicolumn{1}{l}{\text{normal distribution control limits can be used.}}%
\end{array}
$

$%
\begin{tabular}
[c]{c|c|c|c|c|}\cline{2-5}
& \multicolumn{4}{|c|}{$ARL_{0}$}\\\hline
\multicolumn{1}{|c|}{$\zeta$} & $100$ & $250$ & $500$ & $1,000$\\\hline
\multicolumn{1}{|c|}{$0.1$} & 5.995 & 9.041 & 11.743 & 14.485\\\hline
\multicolumn{1}{|c|}{$0.15$} & 5.318 & 7.778 & 9.922 & 12.14\\\hline
\multicolumn{1}{|c|}{$0.2$} & 4.640 & 6.514 & 8.100 & 9.796\\\hline
\multicolumn{1}{|c|}{$0.25$} & 4.186 & 5.816 & 7.208 & 8.607\\\hline
\multicolumn{1}{|c|}{$0.3$} & 3.731 & 5.118 & 6.315 & 7.417\\\hline
\multicolumn{1}{|c|}{$0.35$} & 3.410 & 4.661 & 5.698 & 6.685\\\hline
\multicolumn{1}{|c|}{$0.4$} & 3.089 & 4.204 & 5.080 & 5.952\\\hline
\multicolumn{1}{|c|}{$0.45$} & 2.829 & 3.863 & 4.665 & 5.458\\\hline
\multicolumn{1}{|c|}{$0.5$} & 2.568 & 3.521 & 4.249 & 4.964\\\hline
\end{tabular}
\ $
\end{center}

\subsection{ARL predicted by the heuristic (Section 3.2)}

\begin{center}%
\begin{tabular}
[c]{c}%
Table S2.1\\
\multicolumn{1}{l}{$W$-CUSUM ARL approximations in normal and $t_{3}$}\\
\multicolumn{1}{l}{distributions. $ARL_{0}=500$; changepoint $\tau=100$.}%
\end{tabular}

$%
\begin{tabular}
[c]{c|c|c|c|c|c|c|c|c|}\cline{2-9}
& \multicolumn{4}{|c|}{{\small normal: }${\small (}\theta_{0}=0.98)$} &
\multicolumn{4}{|c|}{t$_{3}${\small : }${\small (}\theta_{0}=1.37)$}\\\hline
\multicolumn{1}{|c|}{${\small (\zeta,h)}$} &
\multicolumn{2}{|c}{${\small (0.10,12.01)}$} &
\multicolumn{2}{|c}{${\small (0.25,7.25)}$} &
\multicolumn{2}{|c}{${\small (0.15,9.86)}$} &
\multicolumn{2}{|c|}{${\small (0.35,5.66)}$}\\\hline
\multicolumn{1}{|c|}{${\small \delta}$} & ${\small W(\delta)}$ &
${\small N(\theta}_{0}{\small \delta)}$ & ${\small W(\delta)}$ &
${\small N(\theta}_{0}{\small \delta)}$ & ${\small W(\delta)}$ &
${\small N(\theta}_{0}{\small \delta)}$ & ${\small W(\delta)}$ &
${\small N(\theta}_{0}{\small \delta)}$\\\hline
\multicolumn{1}{|c|}{{\small 0.125}} & {\small 126} & {\small 127} &
{\small 161} & {\small 157} & {\small 93} & {\small 93} & {\small 131} &
{\small 122}\\\hline
\multicolumn{1}{|c|}{{\small 0.25}} & {\small 57} & {\small 57} & {\small 70}
& {\small 67} & {\small 38} & {\small 37} & {\small 48} & {\small 45}\\\hline
\multicolumn{1}{|c|}{{\small 0.375}} & {\small 36} & {\small 35} & {\small 37}
& {\small 37} & {\small 23} & {\small 22} & {\small 25} & {\small 23}\\\hline
\multicolumn{1}{|c|}{{\small 0.5}} & {\small 26} & {\small 25} & {\small 25} &
{\small 24} & {\small 17} & {\small 16} & {\small 16} & {\small 15}\\\hline
\multicolumn{1}{|c|}{{\small 0.625}} & {\small 20} & {\small 19} & {\small 18}
& {\small 17} & {\small 14} & {\small 12} & {\small 12} & {\small 11}\\\hline
\multicolumn{1}{|c|}{{\small 0.75}} & {\small 17} & {\small 16} & {\small 14}
& {\small 14} & {\small 11} & {\small 10} & {\small 10} & {\small 8}\\\hline
\multicolumn{1}{|c|}{{\small 1.0}} & {\small 11} & {\small 12} & {\small 11} &
{\small 10} & {\small 9} & {\small 7} & {\small 7} & {\small 6}\\\hline
\multicolumn{1}{|c|}{{\small 1.25}} & {\small 10} & {\small 9} & {\small 8} &
{\small 7} & {\small 8} & {\small 6} & {\small 6} & {\small 5}\\\hline
\multicolumn{1}{|c|}{{\small 1.5}} & {\small 9} & {\small 8} & {\small 7} &
{\small 6} & {\small 7} & {\small 5} & {\small 6} & {\small 4}\\\hline
\end{tabular}
\ \ \ \ \ $
\end{center}

Table 3.2 shows the results when $\tau=0$, that is, when the process is out of
control from the outset and the condition in the heuristic that $\tau$ be
large, is not met. While it is clear in this instance that the CUSUM does have
the ability to detect an initial out-of-control situation, the approximation
tends to underestimate quite substantially the true OOC ARL.\newpage

\begin{center}%
\begin{tabular}
[c]{c}%
Table S2.2\\
\multicolumn{1}{l}{$W$-CUSUM ARL approximations in normal and $t_{3}$}\\
\multicolumn{1}{l}{distributions. $ARL_{0}=500$; changepoint $\tau=0$.}%
\end{tabular}

$%
\begin{tabular}
[c]{c|c|c|c|c|c|c|c|c|}\cline{2-9}
& \multicolumn{4}{|c}{{\small normal: }${\small (}\theta_{0}=0.98)$} &
\multicolumn{4}{|c|}{t$_{3}${\small : }${\small (}\theta_{0}=1.37)$}\\\hline
\multicolumn{1}{|c|}{${\small (\zeta,h)}$} &
\multicolumn{2}{|c}{${\small (0.10,12.01)}$} &
\multicolumn{2}{|c}{${\small (0.25,7.25)}$} &
\multicolumn{2}{|c}{${\small (0.15,9.86)}$} &
\multicolumn{2}{|c|}{${\small (0.35,5.66)}$}\\\hline
\multicolumn{1}{|c|}{${\small \delta}$} & ${\small W(\delta)}$ &
${\small N(\theta}_{0}{\small \delta)}$ & ${\small W(\delta)}$ &
${\small N(\theta}_{0}{\small \delta)}$ & ${\small W(\delta)}$ &
${\small N(\theta}_{0}{\small \delta)}$ & ${\small W(\delta)}$ &
${\small N(\theta}_{0}{\small \delta)}$\\\hline
\multicolumn{1}{|c|}{{\small 0.125}} & {\small 146} & {\small 145} &
{\small 171} & {\small 167} & {\small 107} & {\small 106} & {\small 138} &
{\small 125}\\\hline
\multicolumn{1}{|c|}{{\small 0.25}} & {\small 72} & {\small 68} & {\small 78}
& {\small 72} & {\small 48} & {\small 44} & {\small 57} & {\small 47}\\\hline
\multicolumn{1}{|c|}{{\small 0.375}} & {\small 46} & {\small 43} & {\small 46}
& {\small 41} & {\small 32} & {\small 27} & {\small 33} & {\small 25}\\\hline
\multicolumn{1}{|c|}{{\small 0.5}} & {\small 35} & {\small 31} & {\small 32} &
{\small 27} & {\small 25} & {\small 19} & {\small 23} & {\small 16}\\\hline
\multicolumn{1}{|c|}{{\small 0.625}} & {\small 29} & {\small 24} & {\small 25}
& {\small 20} & {\small 21} & {\small 15} & {\small 19} & {\small 12}\\\hline
\multicolumn{1}{|c|}{{\small 0.75}} & {\small 25} & {\small 20} & {\small 21}
& {\small 16} & {\small 19} & {\small 12} & {\small 16} & {\small 9}\\\hline
\multicolumn{1}{|c|}{{\small 1.0}} & {\small 21} & {\small 15} & {\small 16} &
{\small 11} & {\small 16} & {\small 9} & {\small 13} & {\small 7}\\\hline
\multicolumn{1}{|c|}{{\small 1.25}} & {\small 18} & {\small 12} & {\small 14}
& {\small 9} & {\small 15} & {\small 7} & {\small 12} & {\small 5}\\\hline
\multicolumn{1}{|c|}{{\small 1.5}} & {\small 17} & {\small 10} & {\small 13} &
{\small 7} & {\small 14} & {\small 6} & {\small 11} & {\small 4}\\\hline
\end{tabular}
\ \ \ \ $
\end{center}

\subsection{$VdW$-CUSUM. (Section 4)}

\begin{center}
$%
\begin{array}
[c]{c}%
\text{Table\ S4.1}\\
\multicolumn{1}{l}{d_{\delta}\text{ at }ARL_{0}=1,000}%
\end{array}
$%

\begin{tabular}
[c]{c|c|c|c|c|c|c|}\cline{2-7}
& \multicolumn{2}{|c}{$\tau=0$} & \multicolumn{2}{|c}{$\tau=50$} &
\multicolumn{2}{|c|}{$\tau=100$}\\\hline
\multicolumn{1}{|c|}{$\delta$} & $\delta_{1}=0.5$ & $\delta_{1}=1.0$ &
$\delta_{1}=0.5$ & $\delta_{1}=1.0$ & $\delta_{1}=0.5$ & $\delta_{1}%
=1.0$\\\hline
\multicolumn{1}{|c|}{0.25} & 11 & 38 & 4 & 24 & 3 & 18\\\hline
\multicolumn{1}{|c|}{0.4} & 9 & 31 & 3 & 11 & 2 & 10\\\hline
\multicolumn{1}{|c|}{0.5} & \textbf{8} & 24 & \textbf{2} & 8 & \textbf{2} &
5\\\hline
\multicolumn{1}{|c|}{0.75} & \textbf{8} & 15 & \textbf{1} & 3 & \textbf{1} &
2\\\hline
\multicolumn{1}{|c|}{1.0} & \textbf{8} & \textbf{13} & \textbf{1} & \textbf{1}
& \textbf{1} & \textbf{1}\\\hline
\multicolumn{1}{|c|}{1.25} & \textbf{9} & \textbf{12} & \textbf{1} &
\textbf{1} & \textbf{1} & \textbf{1}\\\hline
\multicolumn{1}{|c|}{1.5} & \textbf{9} & \textbf{12} & \textbf{1} & \textbf{1}
& \textbf{1} & \textbf{1}\\\hline
\end{tabular}
\newpage
\end{center}

\subsection{$W^{2}$-CUSUM (Section 6)}

\begin{center}
$%
\begin{array}
[c]{c}%
\text{Table S6}\\
\text{Control limits for the }W^{2}\text{-CUSUM.}%
\end{array}
$

$%
\begin{tabular}
[c]{c|c|c|c|c|c|}\cline{2-6}
& \multicolumn{5}{|c|}{$ARL_{0}$}\\\hline
\multicolumn{1}{|c|}{$\zeta$} & $100$ & $250$ & $500$ & $1,000$ &
$2,000$\\\hline
\multicolumn{1}{|c|}{$0.05$} & 6.57 & 10.08 & 13.39 & 17.34 & 21.61\\\hline
\multicolumn{1}{|c|}{$0.1$} & 5.69 & 8.20 & 10.47 & 12.90 & 15.60\\\hline
\multicolumn{1}{|c|}{$0.15$} & 4.97 & 6.98 & 8.68 & 10.49 & 12.36\\\hline
\multicolumn{1}{|c|}{$0.2$} & 4.40 & 6.08 & 7.45 & 8.87 & 10.29\\\hline
\multicolumn{1}{|c|}{$0.25$} & 3.96 & 5.39 & 6.53 & 7.77 & 8.83\\\hline
\multicolumn{1}{|c|}{$0.3$} & 3.63 & 4.86 & 5.83 & 6.83 & 7.86\\\hline
\multicolumn{1}{|c|}{$0.35$} & 3.28 & 4.39 & 5.25 & 6.11 & 6.97\\\hline
\multicolumn{1}{|c|}{$0.4$} & 3.02 & 4.02 & 4.76 & 5.52 & 6.31\\\hline
\end{tabular}
\ $
\end{center}

\end{document}